\documentclass{aipproc}
\layoutstyle{6x9}
\SetInternalRegister\hbadness{8000} 

\newcommand{\be}{\begin{equation}}
\newcommand{\ee}{\end{equation}}

\newcommand{\beqa}{\begin{eqnarray}}
\newcommand{\eeqa}{\end{eqnarray}}
\newcommand{\nn}{\nonumber}

\newcommand{\spa}{\ \ \ }
\newcommand{\eqref}[1]{(\ref{#1})}
\newcommand{\tr}{\mathop{{\rm Tr}}}
\newcommand{\bo}{\mathbf}
\newcommand{\conus}{\cos (\omega t)}
\newcommand{\sinus}{\sin (\omega t)}

\newcommand{\sn}{\mathop{{\rm sn}}}

\def\boxit#1{\vbox{\hrule\hbox{\vrule\kern8pt
\vbox{\hbox{\kern8pt}\hbox{\vbox{#1}}\hbox{\kern8pt}}
\kern8pt\vrule}\hrule}}
\def\mathboxit#1{\vbox{\hrule\hbox{\vrule\kern8pt\vbox{\kern8pt
\hbox{$\displaystyle #1$}\kern8pt}\kern8pt\vrule}\hrule}}

\def\IB{\relax\hbox{$\inbar\kern-.3em{\rm B}$}}
\def\IC{\relax\hbox{$\inbar\kern-.3em{\rm C}$}}
\def\ID{\relax\hbox{$\inbar\kern-.3em{\rm D}$}}
\def\IE{\relax\hbox{$\inbar\kern-.3em{\rm E}$}}
\def\IF{\relax\hbox{$\inbar\kern-.3em{\rm F}$}}
\def\IG{\relax\hbox{$\inbar\kern-.3em{\rm G}$}}
\def\IGa{\relax\hbox{${\rm I}\kern-.18em\Gamma$}}
\def\IH{\relax{\rm I\kern-.18em H}}
\def\IK{\relax{\rm I\kern-.18em K}}
\def\IL{\relax{\rm I\kern-.18em L}}
\def\IP{\relax{\rm I\kern-.18em P}}
\def\IR{\relax{\rm I\kern-.18em R}}
\def\IZ{\relax\ifmmode\mathchoice
{\hbox{\cmss Z\kern-.4em Z}}{\hbox{\cmss Z\kern-.4em Z}}
{\lower.9pt\hbox{\cmsss Z\kern-.4em Z}}
{\lower1.2pt\hbox{\cmsss Z\kern-.4em Z}}\else{\cmss Z\kern-.4em
Z}\fi}

\def\II{\relax{\rm I\kern-.18em I}}

\def\CN {{\cal N}}

\begin{document}

\title 
      [Title in short, say, Template]
      {D0-branes with non-zero angular momentum}


\author{George K. Savvidy}{
  address={National Research Center Demokritos,
Ag. Paraskevi, GR-15310 Athens, Hellenic Republic},
  email={email:savvidy@argo.nuclear.demokritos.gr}, 
  thanks={This work was commissioned by the AIP}%
}


\copyrightyear  {2001}

\begin{abstract}

In my talk I shall consider the mechanism of {\it self-expansion}
of a system of N D0-branes
into high-dimensional non-commutative world-volume 
investigated by Harmark and Savvidy in \cite{Harmark:2000na}. Here 
D2-brane is formed due to the internal angular momentum
of D0-brane system. The idea is that attractive force of tension
should be cancelled by the centrifugal motion
preventing a D-brane system from collapse to a lower-dimensional one.
I shall also present a new extended solution where 
a total of 9 space dimensions is used to embed a D0-brane system.
In the last section, by performing linear analysis,
the stability of the system is demonstrated.

\end{abstract}
\date{\today}

\maketitle

\section{Introduction}

In the last years there has been increasing interest 
in dimensionally reduced supersymmetric 
Yang-Mills theories \cite{Flume:1985mn,Claudson:1985th,Savvidy:1985gi}. 
One of the reasons is that the reduction of ten-dimensional theory 
to $p+1$ dimension is relevant for the description of Dp-branes,  
$p$-dimensional extended objects carrying 
Ramond-Ramond (RR) charges in type II
superstring theories \cite{Polchinski:1996na}. 
In the extreme reduction to zero 
dimensions it is believed to describe  $D0$-branes, fundamental 
pointlike objects in type IIA superstring theory 
\cite{Witten:1996im}. In certain energy regimes the
dynamics of $N$ such particles can be described by 
the supersymmetric quantum mechanics of $N \times N$
Hermitian matrices obtained from  dimensional reduction of 
$\CN=1$, $D=10$ super-Yang-Mills theory down to $0+1$ dimensions
\footnote{It was also proposed that $0+0$- dimensional matrix model should give 
a Poincare invariant non-perturbative definition of IIB superstring theory,
the so-called IKKT model \cite{Ishibashi:1996jj}.} .

It is believed that supersymmetric 
quantum mechanics of many $D0$-branes in type IIA superstring 
theory is equivalent to a partonic description of
light-front M-theory \cite{Banks:1997vh}, a more 
fundamental underlying M-theory 
\cite{Hull:1995ys,Townsend:1995kk,Witten:1995ex}. 
The existence of matrix formulation of $M$-theory 
\cite{Banks:1997vh}, the BRSS-conjecture,  
crucially relies on the existence within
type-$IIA$ string theory of a tower of massive
BPS particles electrically charged with
respect to the RR $1$-form. These particles 
originally described as black holes in $IIA$ supergravity 
were identified with $D0$-branes and
can be interpreted as Kaluza-Klein particles 
of eleven-dimensional $M$-theory compactified on a circle.
The existence of the $M$-theoretic
Kaluza-Klein tower of states is equivalent to the
statement that supersymmetric Yang-Mills quantum 
mechanics has exactly one bound state for each $N$
\cite{Yi:1997eg,Sethi:1998pa,Green:1998tn}. 

Remarkable aspect of  matrix theory is that not only classical  
gravitational interactions can be produced in the large N-limit, but 
also the appearance of the superstring extended objects 
in terms of pointlike fundamental degrees of freedom. 
One example of such phenomena is the observation that 
a IIA superstring and a system of N D0-branes  can be blown-up 
to a D2-brane by placing it in background field \cite{Emparan:1998rt,Myers:1999ps}.

In my talk I shall consider another mechanism of {\it self-expansion}
of a system of N D0-branes
into high dimensional non-commutative world-volume 
investigated by Harmark and Savvidy in \cite{Harmark:2000na}, here 
D2-brane is formed due to the internal angular momentum
of D0-brane system. The idea is that attractive force of tension
should be cancelled by the centrifugal motion
preventing  a D-brane system from collapse to a lower-dimensional one
\footnote{Similar phenomena appear in the cases of rotating branes on spheres
\cite{McGreevy:2000cw,Grisaru:2000zn}, "giant gravitons", and in the cases
when angular momentum is generated by crossed electric and magnetic BI
fields \cite{Mateos:2001qs}. }. 
I shall also present a new extended solution where 
a total of 9 space dimensions is used to embed a D0-brane system.
In the last section, by performing linear analysis,
the stability of the system is demonstrated.

In the $D$-brane formulation  the dynamics of $N$ D0-branes 
can be described by the supersymmetric quantum mechanics of $N \times N$
Hermitian matrices obtained from  dimensional reduction of $\CN=1$, $D=10$
super-Yang-Mills  theory to 0+1 dimensions 
\cite{Witten:1996im,Myers:1999ps,Taylor:1999gq,Taylor:2000pr}  
(the quantum mechanical model was originally studied in 
\cite{Savvidy:1985gi,Baseian:1979zx,Matinian:1981dj}). 
The effective action of $N$ D0-branes 
is the non-abelian SU(N) Yang-Mills action
plus the Chern-Simons action 
\begin{equation}
\label{BIaction}
S_{YM} = - T_0 (2\pi l_s^2)^2 \int dt \, 
 \tr \left(\frac{1}{4} F_{\mu \nu} F^{\mu \nu} \right)~~,
\end{equation}
where $F_{\mu \nu}$ is the non-abelian $SU(N)$ field strength
in the adjoint representation and \( T_0 = ( g_s l_s )^{-1} \)
is the D0-brane mass.
To write this action in terms of coordinate matrices \( X^i \),
one has to use the dictionary \cite{Harmark:2000na}
\begin{equation}
A_i = \frac{1}{2\pi l_s^2} X^i  ,\spa
F_{0i} = \frac{1}{2\pi l_s^2} \dot{X}^i  ,\spa
F_{ij} = \frac{-i}{(2\pi l_s^2)^2} \, [X^i,X^j]
\label{eq:dict}
\end{equation}
with \( i,j = 1,2,...,9 \) and in $A_0 =0$ gauge we have
\begin{equation}
\label{ham}
S_{YM} = T_0 \int dt \, \tr \left(
\frac{1}{2} \dot{X}^i \dot{X}^i
- \frac{1}{4} \frac{1}{(2\pi l_s^2)^2 } [X^i,X^j][X^i,X^j] \right).
\end{equation}
The equations of motion are
\be
\label{eq:eom}
\ddot{\bo{X}}^i = -  \frac{1}{(2\pi l_s^2)^2 }  \left[ \bo{X}^j \, 
,\left[ \bo{X}^j,\bo{X}^i \right] \, \right]
\ee
and should be taken together with the Gauss constraint 
\begin{equation}
\label{eq:gauss}
\left[\dot{\bo{X}}^i,\bo{X}^i \right] = 0.
\end{equation}
The Chern-Simons action derived in \cite{Myers:1999ps,Taylor:1999gq,Taylor:2000pr}
for the coupling of $N$ D0-branes to bulk RR $C^{(1)}$ and $C^{(3)}$ fields is 
\begin{equation}
\label{CSaction}
S_{CS} = T_0 \int dt \tr \left( C_0 + C_i \dot{X}^i 
+ \frac{1}{2\pi l_s^2 } \left[ C^{(3)}_{0ij}[X^i,X^j]
+ C^{(3)}_{ijk}[X^i,X^j]\dot{X}^k \right] \right)
\end{equation}
and describes the interaction of $N$ D0-branes with
slowly varying background fields of Type IIA supergravity. 
Even though the D0-brane 
world-volume is only one-dimensional a multiple D0-brane system 
can couple to a brane charges of higher dimension.
Myers \cite{Myers:1999ps} considers the system of N D0-branes in
a constant external 4-form RR field strength $F_4 =d C^{(3)}$ and 
has found a stable solution, where the D0-branes are polarized 
and arranged into a static spherical configuration.
A lower-dimensional object under the influence of
higher-form RR fields may nucleate or be 'blown-up' into D2-brane.
The background field imposes an external force
that prevents the collapse of the D-brane to a lower-dimensional one.

Another way to get {\it self-support} against collapse 
is to allow D0-brane system to carry mechanical
angular momentum \cite{Harmark:2000na}.
This new kind of rotating solution of the
system of $N$ D0-branes was constructed by Troels and Konstantin in 
\cite{Harmark:2000na} and in the subsequent articles 
\cite{Savvidy:2000pd,Savvidy:2001td} 
it was demonstrated that this solution describes
a stable D2-brane configuration. 
Below I shall  concentrate mostly on this solution 
and on it generalizations presented in \cite{Savvidy:2001td}.
The basic idea in their construction 
is that the attractive force of tension
should be cancelled by the centrifugal repulsion force.
The earlier work where
membrane solution appeared with non-zero angular momentum 
was \cite{Kikkawa:1986dm}.

\section{D2-brane from multiple rotating D0-branes}

I shall briefly review the spherical D2-brane configuration
of type IIA string theory since a new solution for a rotating system
of $N$ D0-branes presented below
uses the essential elements of this construction.
The solution is equivalent to the spherical membrane solution 
of M(atrix) theory \cite{Banks:1997vh,Kabat:1997im}.

With the aim to construct a membrane with an $S^2$ geometry we
shall embed the $S^2$ in a three-dimensional space spanned by the 
123 directions and consider the ansatz
\begin{equation}
\label{ansatz3}
X_i (t) = \frac{2}{\sqrt{N^2-1}} \bo{L}_i r_i(t) ,\ \ i=1,2,3
\end{equation}
where the $N \times N$ matrices
$\bo{L}_1,\bo{L}_2,\bo{L}_3$ are the generators of the $N$
dimensional irreducible representation of $SU(2)$, with algebra
\begin{equation}
\label{SU2alg}
[\bo{L}_i,\bo{L}_j] = i \, \epsilon_{ijk} \bo{L}_k~~.
\end{equation}
and with the quadratic Casimir
$~\sum_{i=1}^3 \bo{L}_i^2 = \frac{N^2-1}{4}~,~$
so that $~\tr (\bo{L}_i^2) = \frac{N(N^2-1)}{12}.~
$
For vanishing background fields the Hamiltonian is~
$
H = \frac{NT_0}{3}  
\left[ \frac{1}{2} \sum_{i=1}^3 \dot{r}_i^2
+ \frac{\alpha^2}{2} 
\Big( r_1^2 r_2^2 + r_1^2 r_3^2 + r_2^2 r_3^2 \Big) \right]~,
$
where we have introduced a convenient parameter 
$\alpha = \frac{2}{\sqrt{N^2-1}} \frac{1}{2\pi l_s^2}$.
This gives the equations of motion
\be
\label{eom3}
\ddot{r}_1 = - \alpha^2  ( r_2^2+r_3^2 ) r_1,~~~
\ddot{r}_2 = - \alpha^2  ( r_1^2+r_3^2 ) r_2,~~~
\ddot{r}_3 = - \alpha^2  ( r_1^2+r_2^2 ) r_3.
\ee
This system is otherwise known as $0+1$ dimensional
classical SU(2) YM  mechanics \cite{Baseian:1979zx,Matinian:1981dj}.
Let us for simplicity take all radii to be equal to each other: $r_1=r_2=r_3=r$.
With this we have from \eqref{ansatz3} the physical radius of the membrane~ 
$
R^2=X_1^2 + X_2^2 + X_3^2 = \mathbf{I} ~ r^2,~ 
$
where $\mathbf{I}$ is the $N\times N$ identity matrix. 
The last formula  shows that the $N$ D0-branes are
constrained to lie on an \( S^2 \) sphere 
of radius $r$. 
The equations of motion \eqref{eom3} in this case reduce to the
equation \( \ddot{r} = - 2 \alpha^2 r^3 \) and  
the solution is 
$r(t) = R_0 \, \sn ( R_0 \alpha t + \phi)$
oscillating between $R_0$ and $-R_0$
\footnote{Clearly the spherical membrane will 
be classically  point-like at the nodes of the {\sl sinus}.
Thus the membrane solution will break
down after a finite amount of time, 
since the classical solution may not be valid at substringy distances,
and possibly decay into a Schwarzschild black hole \cite{Kabat:1997im}.}.  
One can trust this solution if 
$
|r(t)| \ll l_s \sqrt{N}  ,\spa
|\dot{r}(t)|  \ll 1    ,\spa
|\ddot{r}(t)| \ll l_s^{-1}. 
$
Since we also require \( |r(t)| \gg l_s \) we must have $N \gg 1$.
Thus it is necessary in order to have a large amount of D0-branes to
build a macroscopic spherical membrane.

In order to construct the  rotating ellipsoidal membrane,
viewed as a non-commutative collection of moving D0-branes we shall
take previous configuration of the non-commutative fuzzy sphere
in the 135 directions, and set it to rotate in the transverse space
along three different axis, \textit{i.e.} in the 12, 34 and 56 planes.
We thus use a total of 6 space dimensions to embed our D0-brane system.
The corresponding ansatz is   \cite{Harmark:2000na}
\begin{eqnarray}
\label{ansatz6}
&& 
X_1(t) = \frac{2}{\sqrt{N^2-1}}\, \bo{L}_1\, r_1(t) \spa,~\spa
X_2(t) = \frac{2}{\sqrt{N^2-1}}\, \bo{L}_1\, r_2(t) \spa,
\nn \\ &&
X_3(t) = \frac{2}{\sqrt{N^2-1}}\, \bo{L}_2\, r_3(t) \spa,~\spa
X_4(t) = \frac{2}{\sqrt{N^2-1}}\, \bo{L}_2\, r_4(t) \spa,
\nn \\ &&
X_5(t) = \frac{2}{\sqrt{N^2-1}}\, \bo{L}_3\, r_5(t) \spa,~\spa
X_6(t) = \frac{2}{\sqrt{N^2-1}}\, \bo{L}_3\, r_6(t) \spa.
\end{eqnarray}
where the $N \times N$ matrices
$\bo{L}_1,\bo{L}_2,\bo{L}_3$ are the generators of 
the $N$-dimensional irreducible representation of $SU(2)$.
In this ansatz the matrix structure is 
such that the coordinate matrices are proportional
to the $SU(2)$ generators in pairs and the 
Gauss constraint (\ref{eq:gauss}) is identically satisfied.
It is also the only finite-dimensional
subalgebra of the group of diffeomorphisms of $S^2$, the Sdiff$(S^2)$
\cite{Arakelian:1989zy}.
That is why the $SU(2)$ ansatz is in some sense unique:
it is the only type of solution that carries  over to the 
supermembrane without modification \cite{Taylor:1997dy}.

Substituting the ansatz into \eqref{ham} gives the Hamiltonian
$
H  =  \frac{NT_0}{3} 
\left( \frac{1}{2} \sum_{i=1}^6 \dot{r}_i^2
+ \frac{\alpha^2}{2}\right. 
\Big[ (r_1^2+r_2^2)\, (r_3^2+r_4^2)
\left.
+(r_1^2+r_2^2)\, (r_5^2+r_6^2) 
+ (r_3^2+r_4^2)\, (r_5^2+r_6^2) \Big]\vphantom{\sum_{1}^6}~
\right)~
$
and the corresponding equations of motion 
\begin{eqnarray}
\label{eom6}
&&
\ddot{r}_1 = - \alpha^2\,  ( r_3^2+r_4^2+r_5^2+r_6^2 )\, r_1~,\spa
\ddot{r}_2 = - \alpha^2\,  ( r_3^2+r_4^2+r_5^2+r_6^2 )\, r_2~,
\nn \\ && 
\ddot{r}_3 = - \alpha^2\,  ( r_1^2+r_2^2+r_5^2+r_6^2 )\, r_3~,\spa 
\ddot{r}_4 = - \alpha^2\,  ( r_1^2+r_2^2+r_5^2+r_6^2 )\, r_4~, 
\nn \\ &&
\ddot{r}_5 = - \alpha^2\,  ( r_1^2+r_2^2+r_3^2+r_4^2 )\, r_5~,\spa
\ddot{r}_6 = - \alpha^2\,  ( r_1^2+r_2^2+r_3^2+r_4^2 )\, r_6~. 
\end{eqnarray}
The special solution of these equations,
describing a rotating ellipsoidal membrane with 
three distinct principal radii $R_1$, $R_2$ and $R_3$ is
\cite{Harmark:2000na}
\begin{eqnarray}
\label{solutions}
&&
r_1(t) = R_1 \cos( \omega_1 t + \phi_1 )~,\ \ 
r_2(t) = R_1 \sin( \omega_1 t + \phi_1 )~,
\nn \\ &&
r_3(t) = R_2 \cos( \omega_2 t + \phi_2 )~,\ \ 
r_4(t) = R_2 \sin( \omega_2 t + \phi_2 )~,
\nn \\ &&
r_5(t) = R_3 \cos( \omega_3 t + \phi_3 )~,\ \ 
r_6(t) = R_3 \sin( \omega_3 t + \phi_3 )~.
\end{eqnarray}
This particular functional form of the solution ensures that the
highly non-linear equations for any of the components $r_i$ are
reduced to a harmonic oscillator.  The solution \eqref{solutions}
keeps $r_1^2+r_2^2=R_1^2$ , $~r_3^2+r_4^2=R_2^2~$ and
$~r_5^2+r_6^2=R_3^2$ fixed, which 
allows us to say that the object described by \eqref{solutions}
rotates in six spatial dimensions as a whole without changing its basic
shape. At any point in time one can always choose a coordinate
system in which the object spans only three space dimensions.

Using the equations of motion \eqref{eom6}, the three
angular velocities are determined by the radii, and
do not necessarily have to coincide: 
%
$\omega_1 = \alpha \sqrt{ R_2^2 + R_3^2 }~,\spa
\omega_2 = \alpha \sqrt{ R_1^2 + R_3^2 }~,\spa
\omega_3 = \alpha \sqrt{ R_1^2 + R_2^2 }~.\spa$
%
This dependence of the angular frequency on the radii 
is such that the repulsive force of rotation 
has to be balanced with the attractive force of tension
in order for \eqref{solutions} to be a solution. 
Thus the radii $R_1$, $R_2$ and $R_3$ parameterize \eqref{solutions} 
along with the three phases
$\phi_i$, to produce altogether a six-parameter family of solutions.
In order to exhibit the properties of the solution \eqref{solutions}
one shall  evaluate the energy, 
the components of angular momentum and to find out nonzero D-brane currents.
The non-zero components of $M_{ij}$
are $M_{12}=-M_{21}$, $M_{34}=-M_{43}$ and $M_{56}=-M_{65}$
and correspond to rotations in
the 12, 34 and 56 planes respectively. Their values fit with the
interpretation of the solution as $N$ D0-branes rotating as an ellipsoidal
membrane in that they are time-independent due to conservation law and
proportional to \( N \): ~~~
$M_{12} = \frac{1}{3}N T_0 \omega_1 R_1^2 ~~,~~~
M_{34} = \frac{1}{3}N T_0 \omega_2 R_2^2 ~~,~~~
M_{56} = \frac{1}{3}N T_0 \omega_3 R_3^2 ~~,
$ thus
\begin{equation}
\label{totmoment}
M^2 = \frac{1}{9}(N T_0)^2 \left( \omega_1^2 R_1^4 +
    \omega_2^2 R_2^4 + \omega_3^2 R_3^4 \right),~~~~
    E = \frac{NT_0}{4} \left( \omega_1^2 R_1^2 +
    \omega_2^2 R_2^2 + \omega_3^2 R_3^2 \right).
\end{equation}
Let us now compute the coupling of the solution (\ref{solutions}) 
to the $C^{(3)}$ RR potential. The interaction with the  $C^{(3)}$-field 
is governed by the action (\ref{CSaction}). We denote the 
corresponding current by $J^{ijk}$ 
\begin{equation}
  \label{eq:j}
  J^{ijk} = \frac{1}{2\pi l_s^2 } \tr \left( [X^{[i},X^j]\dot{X}^{k]} \right)~~,
\end{equation}
and we have to impose anti-symmetrization with respect to $ijk$ indices.
For our solution the non-zero components  of Q are:
$  Q_{135}  = {1\over 3}N  \alpha R_1 R_2 R_3 \,
\cos \omega_1 t \cos \omega_2 t \cos \omega_3 t,~$
together with \( Q_{246}~, Q_{146}~,~ Q_{136}~,~Q_{235}~,~Q_{236}~,~Q_{145}~,~Q_{245} \).
From this it is easy to obtain the corresponding $J$'s by
differentiation with respect to time.
{\it Thus the Chern-Simons action \eqref{CSaction} 
shows that the coupling of this system to \( F_{0123} \) is non-vanishing 
and that the spherical membrane solution has a D2-brane dipole
moment.} The  higher currents 
are equal to zero and our system does not carry D4-D8-charges.

In addition I shall present "breathing" brane solutions. 
For that it is convenient to introduce polar coordinates 
$
(\rho ~\cos \phi,~\rho ~\sin \phi)
$
so that the Hamiltonian $\tilde{H}={3H \over NT_0}$ takes the form:
$\tilde{H}= {1 \over 2} \sum^{3}_{i=1} \left[\dot{\rho}^{2}_{i} + \rho^{2}_{i} \dot{\phi}^{2} \right]
 + {1 \over 2} [\rho^{2}_{1}\rho^{2}_{2} + \rho^{2}_{2}\rho^{2}_{3} + 
 \rho^{2}_{3}\rho^{2}_{1}].
$
The conservation integrals are:
$
\rho^{2}_{1}\dot{\phi}_{1} = \tilde{M}_{1},~~~\rho^{2}_{2}\dot{\phi}_{2} = 
\tilde{M}_{2},~~~\rho^{2}_{3}\dot{\phi}_{3} = \tilde{M}_{3},
$
where $\tilde{M}_i ={3M_i \over NT_0}$ and the Hamiltonian takes the form: 
$
\tilde{H} = {1 \over 2} \sum^{3}_{i=1} 
\left[ \dot{\rho}^{2}_{i} + {\tilde{M}^{2}_{i} \over \rho^{2}_{i}} \right] +
{1 \over 2} [\rho^{2}_{1}\rho^{2}_{2} + \rho^{2}_{2}\rho^{2}_{3} + \rho^{2}_{3}\rho^{2}_{1}].
$
The equations of motion are :
\be
\label{newh}
 \ddot{\rho_{i}} = -\rho_{i} (\vec{\rho}^{2}-\rho^{2}_{i} ) 
+ \tilde{M}^{2}_{i} / \rho^{3}_{i} ,~~i=1,2,3
\ee
and our previous solution (\ref{solutions}) is $\rho_{i} = R_{i} = Const, i=1,2,3$ 
and $\dot{\phi}^{2}_{i}
=\omega^{2}_{i} = \tilde{M}^{2}_{i}/R^{4}_{i}= R^{2}_{i+1} + R^{2}_{i+2}$.
In the special case when all coordinates are equal to each other
$\rho_{1} =\rho_{2} =\rho_{3} =\rho(t) $ and can depend on time, the 
"breathing" brane solution, then 
$
\tilde{H}= {3 \over 2}[ \dot{\rho}^{2} + {\tilde{M}^2 \over \rho^{2}}  + \rho^4 ]
$
and corresponding equation can be integrated. The new solution is elliptic function
$\rho = \rho(t)$ \cite{Savvidy:2000pd}
\footnote{A somewhat similar ansatz was proposed in  \cite{Rey:1997iq}, 
where some of the features of the solution were foreseen.}.

In order to increase the number of parameters of the  rotating N D0-brane 
system,
I shall take previous configuration and set it to rotate in the transverse spaces
along three different axis, \textit{i.e.} in the 123, 456 and 789 planes.
Thus I shall use a total of 9 space dimensions to embed D0-brane system.
The corresponding ansatz is   
\begin{eqnarray}
\label{ansatz9}
&& 
X_1(t) = \frac{2}{\sqrt{N^2-1}}\, \bo{L}_1\, r_1(t) \spa,~\spa
X_2(t) = \frac{2}{\sqrt{N^2-1}}\, \bo{L}_1\, r_2(t) \spa,
X_3(t) = \frac{2}{\sqrt{N^2-1}}\, \bo{L}_1\, r_3(t) \spa,
\nn \\ &&
X_4(t) = \frac{2}{\sqrt{N^2-1}}\, \bo{L}_2\, r_4(t) \spa,~\spa
X_5(t) = \frac{2}{\sqrt{N^2-1}}\, \bo{L}_2\, r_5(t) \spa,
X_6(t) = \frac{2}{\sqrt{N^2-1}}\, \bo{L}_2\, r_6(t) \spa,
\nn \\ &&
X_7(t) = \frac{2}{\sqrt{N^2-1}}\, \bo{L}_3\, r_7(t) \spa,~\spa
X_8(t) = \frac{2}{\sqrt{N^2-1}}\, \bo{L}_3\, r_8(t) \spa,
X_9(t) = \frac{2}{\sqrt{N^2-1}}\, \bo{L}_3\, r_9(t) \spa.
\end{eqnarray}
The coordinate matrices are again proportional
to the $SU(2)$ generators and  the 
Gauss constraint (\ref{eq:gauss}) is identically satisfied.
Substituting the last ansatz into \eqref{ham} gives the Hamiltonian:~
$H  =  \frac{NT_0}{3} 
\left( \frac{1}{2} \sum_{i=1}^9 \dot{r}_i^2
+ \frac{\alpha^2}{2}\right. 
\Big[ (r_1^2+r_2^2 +r_3^2)\, (r_4^2+r_5^2 + r_6^2)
\left.
+(r_4^2+r_5^2 + r_6^2)\, (r_7^2+r_8^2 +r_9^2) 
+ (r_7^2+r_8^2 +r_9^2)\, (r_1^2+r_2^2 +r_3^2) \Big]\vphantom{\sum_{1}^6}~
\right)~~
$
and the  equations 
\begin{eqnarray}
\label{eom9}
&&
\ddot{r}_i = - \alpha^2\,  ( r_4^2+r_5^2+r_6^2+r_7^2+r_8^2+r_9^2 )\, r_i~,\spa
i=1,2,3
\nn \\ && 
\ddot{r}_j = - \alpha^2\,  ( r_1^2+r_2^2+r_3^2+r_7^2 +r_8^2+r_9^2)\, r_j~,\spa 
j=4,5,6~, 
\nn \\ &&
\ddot{r}_5 = - \alpha^2\,  ( r_1^2+r_2^2+r_3^2+r_4^2+r_5^2+r_6^2 )\, r_k~,\spa
k=7,8,9~. 
\end{eqnarray}
The special solution of these equations, which ensures that the
highly non-linear equations for any of the components $r_i$ are
reduced to a harmonic oscillator, is:
\begin{eqnarray}
\label{solutions9}
&&
r_1(t) = R_1 \cos( \omega_1 t + \phi_1 )~,\ \ 
r_2(t) = R_1 \sin(\theta_1) \cdot  \sin( \omega_1 t + \phi_1 )~,
r_3(t) = R_1 \cos(\theta_1) \cdot  \sin( \omega_1 t + \phi_1 )~,
\nn \\ &&
r_4(t) = R_2 \cos( \omega_2 t + \phi_2 )~,\ \ 
r_5(t) = R_2 \sin(\theta_2) \cdot \sin( \omega_2 t + \phi_2 )~,
r_6(t) = R_2 \cos(\theta_2) \cdot \sin( \omega_2 t + \phi_2 )~,
\nn \\ &&
r_7(t) = R_3 \cos( \omega_3 t + \phi_3 )~,\ \ 
r_8(t) = R_3 \sin(\theta_3) \cdot \sin( \omega_3 t + \phi_3 )~,
r_9(t) = R_3 \cos(\theta_3) \cdot \sin( \omega_3 t + \phi_3 )~.
\end{eqnarray}
The object described by \eqref{solutions9}
rotates in nine spatial dimensions as a whole without changing its basic
shape. The radii $R_1$, $R_2$ and $R_3$ parameterize \eqref{solutions9} 
along with the six phases $\theta_i$ and $\phi_i$, 
to produce altogether a $nine$-parameter family of solutions.
The energy and the angular momentum are the same \eqref{totmoment}. 
The interaction with the  $C^{(5)}$-field 
is governed by the Chern-Simons action derived in 
\cite{Myers:1999ps,Taylor:1999gq,Taylor:2000pr}
$
{T_0 \over (2\pi l_s^2)^2 }  \int dt S\tr \left(  C^{(5)}_{0ijkl}[X^i,X^j][X^k,X^l]
+ C^{(5)}_{ijklm}[X^i,X^j] [X^k,X^l] \dot{X}^m  \right).
$
We denote the corresponding current by $J^{ijklm}$ 
\begin{equation}
  \label{eq:j5}
  J^{ijklm} = \frac{1}{(2\pi l_s^2)^2 } S\tr \left( [X^{[i},X^j]\dot{X}^{k} [X^{l},X^{m]}]\right)~~.
\end{equation}
One can be convinced that all components  of J are equal to zero.
{\it Thus the Chern-Simons action 
shows that the coupling of this system to \( F_{0ijklm} \) is vanishing 
and that the extended  solution does not carry D4-brane charge.}

\section{Stability analysis  within $SU(2)$ and full $SU(N)$ group}

The purpose of this section is to present a complete stability 
analysis of the fluctuations in the neighborhood 
of the rotating D0-brane solution of \cite{Harmark:2000na}.  
Initially in \cite{Harmark:2000na} were analyzed 
perturbations that do not modify the original $SU(2)$ ansatz. 
In \cite{Savvidy:2001td,Savvidy:2000pd} this
analysis was extended to the case when
perturbations are in the full $SU(N)$ algebra directions.  
In the full $SU(N)$ case there are exactly $N^2+12$ zero-modes, 
of which $N^2-1$
are the consequence of the global color rotation symmetry of the
solution, and $6$ are associated with global space rotations.
All other modes are completely stable and execute harmonic
oscillations around the original trajectory.

Let me present the  stability analysis of the $SU(2)$ 
($l=1$)  perturbation of the system  
 \cite{Harmark:2000na} and then move to more 
general  $SU(N)$ ($l=2,3,4,....$)  perturbations
\cite{Savvidy:2001td,Savvidy:2000pd}. 
The equations of variation which follow from equations in polar coordinates 
(\ref{newh}) are:
\beqa
\delta\ddot{\rho_{1}} = -4 (R^{2}_{2}  + R^{2}_{3})\delta \rho_{1} 
- 2R_{1}R_{2}  \delta \rho_{2} - 2R_{1}R_{3}  \delta \rho_{3} \\
\delta\ddot{\rho_{2}} =  - 2R_{2}R_{1}    \delta \rho_{1} 
-4 (R^{2}_{1}  + R^{2}_{3})\delta \rho_{2} - 2R_{2}R_{3} \delta \rho_{3} \\
\delta\ddot{\rho_{3}} = - 2R_{3}R_{1}\delta \rho_{1} - 2R_{3}R_{2} \delta \rho_{2}
-4 (R^{2}_{1}  + R^{2}_{2})\delta \rho_{3},
\eeqa
and have only positive modes \cite{Savvidy:2000pd}
\beqa
\Omega^{2}_1 = 4(R^{2}_{1}  + R^{2}_{2} + R^{2}_{3}),~~~~~~~~~~~~~~~~~~~~~~~~~~~~~~~~~~~~~~~~~~~\\
\Omega^{2}_{2,3} = 2(R^{2}_{1}  + R^{2}_{2} + R^{2}_{3}) \pm 
\sqrt{2( (R^{2}_{1} - R^{2}_{2} )^{2}  +  (R^{2}_{2} - R^{2}_{3} )^{2}  +(R^{2}_{3} - R^{2}_{1} )^{2})}.
\eeqa
To consider perturbations in all directions of underlying $SU(N)$ group we 
should represent $SU(N)$ generators $\bo{Y}^l_m$
as higher order monomials in the $N \times N$ matrix 
generators $\bo{L}_i,~i=1,2,3$ of the $SU(2)$ group
\be
\label{bases}
\bo{Y}^l_m = \sum_{i_1 ,...,i_l} c^l_{m~(i_1 ,...,i_l)} ~~\bo{L}_{i_1} \cdots \bo{L}_{i_l} ~.
\ee 
The total number of generators is then $\sum_{l=1}^{N-1} (2l+1) = N^2-1$ 
as it should be for $SU(N)$.
The general explicit construction of the $\bo{Y}^l_m$ is due to 
Schwinger (see also \cite{hoppe,Biran:1987ae,Arakelian:1988gm,Floratos:1988yp}).

Let us represent the rotating D0-brane solution  in a more 
convenient form:  
$
\bo{X}^i(t) = \bo{L}_i \, R \, \cos(\omega t),~~
\tilde{\bo{X}}^i(t) = \bo{X}^{i+3}(t) = \bo{L}_i \, R \, \sin(\omega t),~~~i=1,2,3.
$
In what follows we will set $R=1$, with $\omega$ equal 
to $\omega^2 = 2R^2 =2$. In the basis provided 
by the spherical operators $\bo{Y}^l_m$ (\ref{bases}) we have  
\cite{Arakelian:1989zy,Taylor:1997dy}, 
\beqa
\left[\bo{L}_z, \bo{Y}^l_m\right] &=& m \, \bo{\bo{Y}}^l_m 
\qquad\qquad\qquad \textrm{for ~~}l=1,\ldots,N-1\nn \\
\left[\bo{L}_{\pm} ,{\bo{Y}}^l_m \right] &=& \sqrt{ (l \mp m)(l \pm m +1) } \, {\bo{Y}}^l_{m\pm1}.
\label{eq:Ydef}
\eeqa
We will not use the explicit form of these matrices, as the defining
relations \eqref{eq:Ydef} is all that is needed.
The properties under Hermitian conjugation can be summed up as
$
\bo{Y}_{l,m}^{\dagger} = (-1)^m \, \bo{Y}_{l,-m}.~
$
I shall first consider the perturbations which are parallel 
in space to one of the directions of our solution, that is 
in the directions 123456 and then in the directions 789.
Let us decompose the fluctuation fields in the basis defined by 
$\bo{Y}^l_m$, where $l$ runs from $1$ to $N-1$ 
\be
\label{eq:eta-def}
\delta \bo{X}^i = \sum_{m=-l}^l \bo{Y}^l_m \, \xi_m^i ~~,\qquad \delta
\tilde{\bo{X}}^i = \sum_{m=-l}^l \bo{Y}^l_m \, \eta_m^i \qquad
i=1,2,3.  
\ee 
We do not explicitly show an $l$ index on the $\eta,\xi$. 
We should also impose 
$
\xi_m^{*} = (-1)^m  \xi_{-m} ~~~\textrm{and}~~~~~~ \eta_m^{*} = (-1)^m  
\eta_{-m} ~~\textrm{for all}~~m=-l,\ldots,l~~.
$
The variational equations of motion  are
\beqa
\label{eq:var-eom}
- \delta \ddot{\bo{X}}^i &=& 
\left[ \delta \bo{X}^j \, ,\left[ \bo{X}^j,\bo{X}^i \right] \, \right] +
\left[ \bo{X}^j \, ,\left[ \delta \bo{X}^j,\bo{X}^i \right] \, \right] +
\left[ \bo{X}^j \, ,\left[ \bo{X}^j, \delta \bo{X}^i \right] \, \right] + \nn\\
&&\left[ \delta \tilde{\bo{X}}^j \, ,\left[ \tilde{\bo{X}}^j,\bo{X}^i \right] \, \right] +
\left[ \tilde{\bo{X}}^j \, ,\left[ \delta \tilde{\bo{X}}^j,\bo{X}^i \right] \, \right] +
\left[ \tilde{\bo{X}}^j \, ,\left[ \tilde{\bo{X}}^j, \delta \bo{X}^i \right] \, \right] ~~.
\eeqa
The constraint equation looks like
\be
\label{eq:var-constr}
\sum_{i,m} \left[\delta \dot{\bo{X}^i} , \bo{X}^i\right] + 
\left[ \dot{\bo{X}^i} , \delta \bo{X}^i\right] + 
\left[\delta \dot{\tilde{\bo{X}}^i} , \tilde{\bo{X}}^i\right] + 
\left[ \dot{\tilde{\bo{X}}^i} , \delta \tilde{\bo{X}}^i\right]  = 0.
\ee
Using the commutation relations (\ref{eq:Ydef}) we get for the constraint
$
\sum_i \bo{L}^i_{nm} 
( \conus \dot{\xi}_m^i + 
\omega \sinus {\xi}_m^i + 
\sinus \dot{\eta}_m^i - \omega \conus \eta_m^i ) =0~~,
$
where $\bo{L}^i_{nm}$ are now the $SU(2)$ generators in the 
$(2l+1)\times(2l+1)$ representation. 
In the co-moving  coordinates 
\be
\label{eq:comov}
u_m^i= \conus \xi_m^i + \sinus \eta_m^i~~~
\textrm{and}~~~~ v_m^i = -\sinus \xi_m^i +\conus \eta_m^i
\ee

\be
\begin{array}{l}
\textrm{the constraint looks simpler,~~~~~~~~~~~~~~~~~~~~~~~~~~~~~~~~~
}\sum_{i,m} \bo{L}^i_{nm} \left ( \dot{u}_m^i -2 \omega v_m^i \right ) =0~.
\label{constr}
\end{array}
\ee
The variational equation of motion (\ref{eq:var-eom}) after substituting
the fields (\ref{eq:eta-def}) is
\be
\ddot{\xi}_n^i + l(l+1) \,  \xi_n^i = 
 \conus \, \left( \bo{L}^j_{n n^{\prime} }\bo{L}^i_{n^{\prime} m} +
 i \, \epsilon_{jik} \bo{L}^k_{nm} \right) \,
\left( \conus \, \xi_m^j +\sinus \, \eta_m^j \right) ~.               
\ee
The decoupling of the modes with different $l$ is seen to be a direct
consequence of (\ref{eq:Ydef}), and more fundamentally, of the pure $SU(2)$
structure of the original background solution (\ref{solutions}).
The equation for $\eta$ is gotten by exchanging cosines for sines and $\xi$ for $\eta$.
In the co-moving  coordinates (\ref{eq:comov})
the equation becomes a linear system with $constant$ coefficients:
\beqa
\ddot{u}_n^i + \left( l(l+1) -2 \right) u_n^i -2\omega \, \dot{v}_n^i &=& 
\left( \bo{L}^j_{n n^{\prime} }\bo{L}^i_{n^{\prime} m} +
 i \, \epsilon_{jik} \bo{L}^k_{nm} \right) \, u_m^j ~, \label{eq:fnl1}\\
\ddot{v}_n^i + \left( l(l+1) -2 \right) \, v_n^i + 2\omega \, \dot{u}_n^i  &=& 0 ~. 
\label{eq:fnl2}
\eeqa
Thus we shall analyze the system of equations (\ref{constr}) 
(\ref{eq:fnl1}),  (\ref{eq:fnl2}).  
The $rhs$ of \eqref{eq:fnl1} is a matrix acting on a $3(2l+1)$ component vector
and the eigenvalues $\Lambda$ of this block matrix 
are given in the table, together with their multiplicity \cite{Savvidy:2001td}.
Choose  a fixed frequency ansatz
$
u^i_n(t) = e^{ i \Omega t} \, u^i_n ~~,~~~v^i_n(t) = e^{ i \Omega t} \, v^i_n ~~.
$
The second equation \eqref{eq:fnl2} can be solved as
\be
v^i_n = \frac{ -2\sqrt{2} \, i \, \Omega^2} { l(l+1) -2 - \Omega^2} \,  u^i_n ~~,
\ee
and then substituted back into the first equation \eqref{eq:fnl1}.
In the basis in which the matrix on the $rhs$ is diagonalized,
it can be replaced with its respective eigenvalue $\Lambda$,
resulting in an algebraic equation for the $\Omega$: ~
$
\left (l\left (l+1\right )-2-\Omega^2 \right )^{2}-8\,\Omega^2 =\Lambda\,
\left (l\left (l+1\right )-2-\Omega^2 \right )~.
$
Finally, this quadratic equation can be solved, 
$
\Omega_{1,2}^2  = -{1\over 2}\,\Lambda+ l(l+1)+2 
\pm {1\over 2}\,\sqrt {{\Lambda}^{2}-16\,\Lambda+32\, l(l+1) }~
$
and the corresponding modes are given in the table
\be
\label{eq:modes}
\begin{array}{|c|c|c|c|}
\Lambda & \Omega^{2}_{1} & \Omega^{2}_{2} & \textrm{multiplicity} \\
\hline
 l(l+1) -2 &   0              & l^2 + \, l +6    & 2l+1             \\
 2l        &   l^2 -3l +2     & l^2 + 3l +2    & 2l+3           \\
 -(2l+2)   &   l^2 -\, l~     & l^2 + 5l +6    & 2l-1             \\
\hline
\end{array}
\ee
Note that the number of zero modes changes from $9$ for the case $l=1$ and $12$ for $l=2$,
to $2l+1$ for arbitrary $l>2$. 
Thus the total number of zero modes is
the sum $
9 + 12 + \sum_{l=3}^{N-1} (2l+1) = N^2 + 12
$. 
Now I shall consider perturbations that are in the directions 789, 
if we had oriented the original solution along 123456.  
The perturbations
$
\delta\bo{X}^k = \sum_m  \bo{Y}^l_{m} \,  \zeta_m^k ~~~~~\textrm{for}~~~k=7,8,9
$
satisfy the simple harmonic equation
$
\ddot{\zeta}_m^k + l(l+1) \, {\zeta}_m^k = 0 ~~.
$
This clearly has only positive frequencies and is therefore stable. For $l=1$
all the $9$ modes have the same frequency as the original solution, corresponding
to infinitesimal global rotations of the system into the 789 hyperplane. 
The counting goes as follows, there are $9\times 2 =18$ first order degrees of
freedom here, which coincides with the dimensionality of the grassmanian manifold
of embeddings of a 6-hyperplane into ${R}^9$, i.e. $ \frac{SO(9)}{ SO(6) \times SO(3) } $.
From these results it follows that,
zero-modes notwithstanding, all the frequencies in the system are
positive, and arbitrary small perturbation will remain bounded for all
times  \footnote{The same problem was considered also
in the paper \cite{Axenides:2000mn}. However the authors of 
\cite{Axenides:2000mn} initially arrived at the Mathiew equation  
instead of the equations (\ref{eq:fnl1}),  (\ref{eq:fnl2}), (\ref{constr}) 
and therefore to the opposite conclusion, namely 
that there exist solutions of the linearized perturbation equations
which grow exponentially (\cite{Axenides:2000mn},v1). It was found then that an algebraic mistake appeared
in the calculation  so that finally they ended up with the same 
conclusion (\cite{Axenides:2000mn},v2).}.

In conclusion I wish to thank the conference organizers,
especially Tsukasa Tada, Noboru Kawamoto and Hajime Aoki, for their hospitality. 
This work was supported in part by Tohwa University and 
the EEC Grant no. HPRN-CT-1999-00161.



\begin{thebibliography}{99}


\bibitem{Harmark:2000na}
T.~Harmark and K.~G.~Savvidy,
{\it Ramond-Ramond field radiation from rotating ellipsoidal membranes },
Nucl.\ Phys.\ B {\bf 585} (2000) 567
[hep-th/0002157].

\bibitem{Flume:1985mn}
R.~Flume,
{\it On Quantum Mechanics With Extended Supersymmetry 
And Nonabelian Gauge Constraints,}
Annals Phys.\  {\bf 164} (1985) 189.

\bibitem{Claudson:1985th}
M.~Claudson and M.~B.~Halpern,
{\it Nucl.\ Phys.\ B} {\bf 250}, 689 (1985).

\bibitem{Savvidy:1985gi}
G.~K. Savvidy, {\it {{Yang--Mills} Quantum Mechanics}},  {\em Phys. Lett.} {\bf
  B159} (1985) 325.


\bibitem{Polchinski:1996na}
J.~Polchinski,
{\it TASI lectures on D-branes,}
hep-th/9611050.


\bibitem{Witten:1996im}
E.~Witten, {\it {Bound States of Strings and p-Branes}},  {\em Nucl. Phys.}
  {\bf B460} (1996) 335--350.

\bibitem{Ishibashi:1996jj}
N.~Ishibashi, H.~Kawai, Y.~Kitazawa and A.~Tsuchiya,
{\it Nucl.\ Phys.\ B} {\bf 498}, 467 (1997).

\bibitem{Hull:1995ys}
C.~M.~Hull and P.~K.~Townsend,
{\it Unity of superstring dualities,}
Nucl.\ Phys.\ B {\bf 438} (1995) 109.

\bibitem{Townsend:1995kk}
P.~K.~Townsend,
{\it The eleven-dimensional supermembrane revisited,}
Phys.\ Lett.\ B {\bf 350} (1995) 184.

\bibitem{Witten:1995ex}
E.~Witten,
{\it String theory dynamics in various dimensions,}
Nucl.\ Phys.\ B {\bf 443} (1995) 85.

\bibitem{Banks:1997vh}
T.~Banks, W.~Fischler, S.~H. Shenker, and L.~Susskind, 
{\it Phys. Rev.} {\bf D55} (1997) 5112--5128.


\bibitem{Yi:1997eg}
P.~Yi,
{\it Witten index and threshold bound states of D-branes,}
Nucl.\ Phys.\ B {\bf 505}, 307 (1997).

\bibitem{Sethi:1998pa}
S.~Sethi and M.~Stern,
{\it D-brane bound states redux,}
Commun.\ Math.\ Phys.\  {\bf 194}, 675 (1998).

\bibitem{Green:1998tn}
M.~B.~Green and M.~Gutperle,
{\it JHEP {\bf 9801} (1998) 005.}


\bibitem{Emparan:1998rt}
R.~Emparan, {\it {{Born-Infeld} Strings Tunneling to {D-branes}}},  {\em Phys.
  Lett.} {\bf B423} (1998) 71--78.

\bibitem{Myers:1999ps}
R.~C.~Myers,
{\it Dielectric-branes,}
JHEP {\bf 9912} (1999) 022
[hep-th/9910053].

\bibitem{McGreevy:2000cw}
J.~McGreevy, L.~Susskind and N.~Toumbas,
{\it JHEP}{\bf 0006} (2000) 008
[hep-th/0003075].

\bibitem{Grisaru:2000zn}
M.~T.~Grisaru, R.~C.~Myers and O.~Tafjord,
{\it SUSY and Goliath,}
JHEP {\bf 0008} (2000) 040.

\bibitem{Mateos:2001qs}
D.~Mateos and P.~K.~Townsend,
{\it  Supertubes,}
hep-th/0103030.

\bibitem{Taylor:1999gq}
W.~I.~Taylor and M.~Van Raamsdonk,
{\it Nucl.\ Phys.\ B {\bf 558} (1999) 63. }
[hep-th/9904095].


\bibitem{Taylor:2000pr}
W.~I.~Taylor and M.~Van Raamsdonk,
{\it Nucl.\ Phys.\ B} {\bf 573} (2000) 703.


\bibitem{Baseian:1979zx}
G.~Z. Baseian, S.~G. Matinian, and G.~K. Savvidy, {\it {Nonlinear Plane Waves
  In Massless {Yang--Mills} Theory. (In Russian})},  {\em Pisma Zh. Eksp. Teor.
  Fiz.} {\bf 29} (1979) 641.



\bibitem{Matinian:1981dj}
S.~G. Matinian, G.~K. Savvidy, and N.~G. T.-A. Savvidy, {\it {Classical
  {Yang--Mills} Mechanics. Nonlinear Color Oscillations}},  {\em Sov. Phys.
  JETP} {\bf 53} (1981) 421--425.

\bibitem{Savvidy:2000pd}
K.~G.~Savvidy,
{\it The discrete spectrum of the rotating brane},
hep-th/0004113.


\bibitem{Savvidy:2001td}
K.~G.~Savvidy and G.~K.~Savvidy,
{\it Stability of the rotating ellipsoidal D0-brane system},
Phys.\ Lett.\ B {\bf 501}, 283 (2001)
[hep-th/0009029].

\bibitem{Kikkawa:1986dm}
K.~Kikkawa and M.~Yamasaki,
{\it Can The Membrane Be A Unification Model?,}
Prog.\ Theor.\ Phys.\  {\bf 76} (1986) 1379.

\bibitem{Arakelian:1988gm}
T.~A.~Arakelian and G.~K.~Savvidy,
{\it Cocycles Of Area Preserving Diffeomorphisms,
Phys.\ Lett.\ B} {\bf 214}, 350 (1988).

\bibitem{Arakelian:1989zy}
T.~A.~Arakelian and G.~K.~Savvidy,
{\it Geometry Of A Group Of Area Preserving Diffeomorphisms,}
Phys.\ Lett.\ B {\bf 223} (1989) 41.


\bibitem{Kabat:1997im}
D.~Kabat and W.~Taylor, {\it Spherical Membranes in Matrix Theory,  
  Adv.Theor.Math.Phys.} {\bf 2} (1998) 181.


\bibitem{Taylor:1997dy}
W.~I.~Taylor,
 {\it Lectures on D-branes, gauge theory and M(atrices),}
in {\it C97/09/02.4}
hep-th/9801182.

\bibitem{Rey:1997iq}
S.~Rey,
{\it Gravitating M(atrix) Q-balls,}
hep-th/9711081.

\bibitem{hoppe}
J.~Hoppe.
\newblock {Ph.D. Thesis MIT}.
\newblock 1982.

\bibitem{Biran:1987ae}
B.~Biran, E.~G.~F. Floratos, and G.~K. Savvidy, {\it {The Selfdual Closed
  Bosonic Membranes}},  {\em Phys. Lett.} {\bf B198} (1987) 329.

\bibitem{Floratos:1988yp}
E.~G.~Floratos and J.~Iliopoulos,
{\it A Note On The Classical Symmetries Of The Closed Bosonic Membranes,''
Phys.\ Lett.\ B} {\bf 201}, 237 (1988).

\bibitem{Axenides:2000mn}
M.~Axenides, E.~G.~Floratos and L.~Perivolaropoulos,
Metastability of spherical membranes in supermembrane and matrix theory.
hep-th/0007198 v1,v2

\end{thebibliography}
\end{document}